# Machine learning algorithms to infer trait-matching and predict species interactions in ecological networks

Maximilian Pichler[1,*], Virginie Boreux[2], Alexandra-Maria Klein[2], Matthias Schleuning[3], Florian Hartig[1]

[1] Theoretical Ecology, University of Regensburg, Universitätsstraße 31, 93053 Regensburg, Germany

[2] Nature Conservation and Landscape Ecology, University of Freiburg, Tennenbacherstraße 4, 79106 Freiburg, Germany

[3] Senckenberg Biodiversity and Climate Research Centre (SBiK-F), Senckenberganlage 25, 60325 Frankfurt (Main), Germany

[*] corresponding author, maximilian.pichler@biologie.uni-regensburg.de

## Abstract

1. Ecologists have long suspected that species are more likely to interact if their traits match in a particular way. For example, a pollination interaction may be more likely if the proportions of a bee's tongue fit a plant's flower shape. Empirical estimates of the importance of trait-matching for determining species interactions, however, vary significantly among different types of ecological networks.

2. Here, we show that ambiguity among empirical trait-matching studies may have arisen at least in parts from using overly simple statistical models. Using simulated and real data, we contrast conventional generalized linear models (GLM) with more flexible Machine Learning (ML) models (Random Forest, Boosted Regression Trees, Deep Neural Networks, Convolutional Neural Networks, Support Vector Machines, naïve Bayes, and k-Nearest-Neighbor), testing their ability to predict species interactions based on traits, and infer trait combinations causally responsible for species interactions.

3. We find that the best ML models can successfully predict species interactions in plant-pollinator networks, outperforming GLMs by a substantial margin. Our results also demonstrate that ML models can better identify the causally responsible trait-matching combinations than GLMs. In two case studies, the best ML models successfully predicted species interactions in a global plant-pollinator database and inferred ecologically plausible trait-matching rules for a plant-hummingbird network from Costa Rica, without any prior assumptions about the system.

4. We conclude that flexible ML models offer many advantages over traditional regression models for understanding interaction networks. We anticipate that these results extrapolate to other ecological network types. More generally, our results highlight the potential of machine learning and artificial intelligence for inference in ecology, beyond standard tasks such as image or pattern recognition.

# Introduction

The understanding and analysis of species interactions in ecological networks has become a central building block of modern ecology. Research in this field, however, has concentrated in particular on analyzing observed network structures (e.g. González, Dalsgaard, & Olesen, 2010; Poisot, Stouffer, & Gravel, 2015; Galiana et al. 2018; Mora et al. 2018). Our understanding of why particular species interact, and others not, is comparatively less developed (cf. Poisot, Stouffer, & Gravel, 2015; Bartomeus et al. 2016). A key hypothesis regarding this question is that species interact when their functional properties (traits) make an interaction possible (e.g. Jordano, Bascompte, & Olesen, 2003; Eklöf et al. 2013). In plant-pollinator networks, for example, one would imagine that an interaction is easier to achieve when the tongue or body of the bee match with the shape and size of the flower (Stang, Klinkhamer & van der Meijden, 2007; Garibaldi et al. 2015). The idea that interactions will occur when traits are compatible is known as trait-matching (e.g. Schleuning, Fründ & García, 2015, see also Fig. 1).

The assumption that trait-matching is important for species interactions is engraved in many other ecological ideas and hypotheses. For example, trait-matching is a prerequisite for the idea of pollination syndromes (i.e. the hypothesis that flower and pollinator traits co-evolve, Faegri & van der Pjil 1979; see also Fenster et al. 2004, Ollerton et al. 2009, and Rosas-Guerrero et al. 2014). Moreover, it has been suggested that trait-matching occurs also in other mutualistic ecological networks e.g. fruit-frugivore interactions (e.g. Dehling et al. 2014), or antagonistic ecological networks, e.g. host-predator or host-parasitoid networks (Gravel et al. 2013; see also Eklöf et al. 2013). Trait-matching between species has ample consequences for fundamental research, such as the identification and prediction of species interactions (Bartomeus et al. 2016, see Valdovinos, 2019), but also impacts ecosystem management. For example, it could be used for identifying effective pollinators to optimize production of pollinator-dependent crops (Garibaldi et al. 2015; Bailes et al. 2015; see Potts et al. 2016). Finally, explaining and predicting links between interaction partners from information about their properties has applications far beyond ecology. An example is molecular medicine, where analogue concepts are used to study gene association (e.g. Yamanishi et al. 2008; van Laarhoven & Marchiori 2013; Menden et al. 2013; Zhang et al. 2018) or harmful drug-drug interactions (e.g. Tari et al. 2010; Cheng & Zhao 2014).

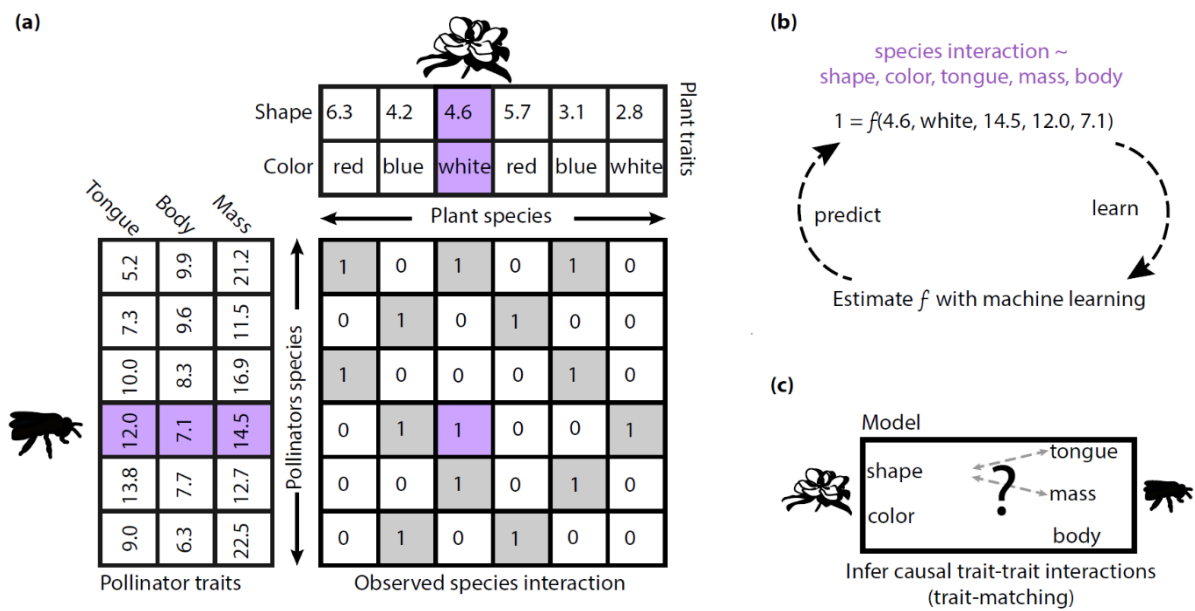

**Figure 1:** An illustration of the trait-matching concept. a) Two classes of organisms, each with their own traits, interact in a bipartite network. b) The goal of the statistical algorithm is to predict the probability of a plant-pollinator interaction, based on their trait values and c) to infer the trait-trait interaction structure (trait-matching) that is causally responsible for those interactions.

While the idea of trait-matching itself is intuitive, it is less clear how important this mechanism is for determining species interactions (Eklöf et al. 2013, Bartomeus et al. 2016). On the one hand, recent findings in plant-pollinator networks support the concept of pollination syndromes (Rosas-Guerrero et al. 2014) and the utility of syndromes for predicting or understanding species interactions (Danieli-Silva et al. 2012; Murúa & Espíndola 2015; Fenster et al. 2015; see Garibaldi et al. 2015). Recent studies also demonstrate that species interactions can be reasonably well predicted with phylogenetic predictors (Pearse & Altermatt 2013; Brousseau, Gravel, & Handa, 2018; Pomeranz et al. 2019), which supports the idea of trait-matching when assuming that traits are phylogenetically conserved. Similarly, studies of mutualistic pollination and seed-dispersal networks have accumulated evidence for strong signals of trait-matching, in particular in diverse tropical ecosystems (Maglianesi et al. 2014; Dehling et al. 2016). On the other hand, many ecological networks show low to moderate levels of specialization (Blüthgen et al. 2007) and high flexibility in partner choice (Bender et al. 2017), questioning the idea of strong co-evolutionary feedback loops between plants and animals (Janzen 1985; Ollerton et al. 2009). Moreover, while there is some direct evidence for trait-trait relationships as predictors for trophic interactions in simple prey-predator networks (Gravel et al. 2013), recent models that relied solely on trait-trait predictors (without phylogenetic predictors) showed only moderate performance in predicting species interactions (Brousseau, Gravel, & Handa, 2018; Pomeranz et al. 2019).

Progress on these questions is complicated by the fact that, until very recently, analyses of empirical networks relied almost exclusively on conventional regression models and phylogenetic predictors (Pearse & Altermatt 2013; Brousseau, Gravel, & Handa, 2018; Pomeranz et al. 2019), or on simple regression trees (e.g. Berlow et al. 2009). Reasonable doubts exist as to whether these models are flexible enough to capture the way traits give rise to interactions (see e.g. Mayfield & Stouffer 2017). Machine Learning (ML) models could be a solution to this problem. Modern ML models can flexibly detect interactions between predictors (trait-trait interactions), depend on fewer a-priori assumptions and usually achieve higher predictive performance than traditional regression techniques (e.g. Breiman 2001b). State-of-the-art deep learning algorithms can detect complex pattern (e.g. LeCun, Bengio, & Hinton, 2015) and excel in tasks such as image or species recognition (e.g. Tabak et al. 2018, Gray et al. 2019). In food webs, recent findings demonstrate the potential of ML models or predicting species interactions. For example, Desjardins-Proulx et al. 2017 report that both a k-nearest-neighbor and random forest (based on phylogenetic relationships and traits) can successfully predict food web interactions. It therefore seems promising to further explore the performance of machine learning algorithms for predicting species interactions from measurable traits, and whether those more flexible models change our view on the importance of trait matching for plant-pollinator interactions.

When assessing the suitability of ML algorithms for this problem, it is important to note that, while ML models tend to excel in predictive performance, their interpretation is often challenging (e.g. Ribeiro, Singh, & Guestrin, 2016). Ecologists, however, would likely not be satisfied with predicting species interactions, but would also want to know which traits are causally responsible for those interactions, for instances due to their importance as essential biodiversity variables (see Kissling et al. 2018). Unlike for statistical models, however, fitted ML models typically provide no direct information about how they generate their predictions. In recent years, also in response to issues such as fairness and discrimination (see Olhede & Wolfe 2018), techniques aiming at interpreting fitted ML models have emerged (e.g. Guidotti et al. 2018). For example, permutation techniques (Fisher, Rudin, & Dominici 2018) allow estimating the importance of predictors for any kind of model, similar to the variable importance in tree-based models (Breiman 2001a). In this case, however, we are not primarily interested in the effects of a single predictor, but we want to know how interactions between predictors (trait-trait-matching) influence interaction probabilities. A suggested solution to this problem is the H-statistic (Friedman & Popescu 2008), which uses partial dependencies to estimate feature-feature (trait-trait) interactions from fitted ML models. Assuming that networks emerge due to a few important trait-trait interactions (Eklöf et al. 2013), the H-statistic should be able to identify those from a fitted ML, but to our knowledge, the efficacy of this or similar techniques in inferring causal traits has not yet been demonstrated.

The purpose of this study is to (i) systematically assess the predictive performance of different ML models for the identification of trait-matching in plant-pollinator networks and (ii) to investigate if causal traits can be extracted from the fitted models with the H-statistics. We consider the most common ML models (k-nearest-neighbor, random forest, boosted regression trees, deep neural networks, support vector machine, naïve Bayes, and convolutional neural networks), with standard generalized linear model (GLM) as a benchmark. We apply all models to simulated and empirical plant-pollinator networks to establish how networks properties influence their predictive performance, and to test if the causally responsible trait-trait interactions be inferred from the fitted models. We ask the following questions: **(1)** Which algorithms display the highest predictive performance for simulated plant-pollinator networks, varying network sizes, observation times, and species abundances? **(2)** Can we retrieve the true underlying trait-trait interaction structure (trait-matching) in the simulated plant-pollinator networks from the fitted ML models? We demonstrate the practical utility of the developed methods by predicting interactions in a global crop-pollinator interaction database, and by inferring the causal trait-trait interaction structure in a Costa Rican plant-hummingbird network.

**Table 1.** Machine learning models and their usage in ecological trait-matching.

| ML models | Type | Design principle | Applied with trait-matching |
|---|---|---|---|
| random forest (RF) | tree-based | Ensemble of a finite number of regression trees (see Breiman, 2001a). | Desjardins-Proulx et al. 2017; Masahiro & Rillig, 2017; Hu et al. 2016 |
| boosted regression trees (BRT) | tree-based | After fitting the first weak regression tree to the response, subsequent regression trees are fitted on the previous residuals (see Friedman, 2001). | He et al. 2017; Rayhan et al. 2017 |
| k-nearest-neighbor (kNN) | distance-based | Given new point X, nearest k neighbors determine response. | Desjardins-Proulx et al. 2017 (as recommender system); Rodgers et al. 2010 |
| support vector machines (SVM) | distance-based | In the n-dimensional feature space, a hyperplane to separate the classes is fitted (see Cristianini & Shawe-Taylor 2000). | Fang et al. 2013 |
| deep neural networks (DNN) | neural networks | By learning to represent the input over several hidden layers, they are able to identify the patterns in the data for the task | Wen et al. 2017 |
| convolutional neural networks (CNN) | neural networks | Topological patterns in the input space (images, sequences) are preserved and processed by a number of kernels to extract features (see LeCun, Bengio, & Hinton, 2015). | Liu et al. 2016 |
| naive Bayes | probabilistic | The model learns the probability | Fang et al. 2013 |

| | classifier | | belonging to a class given a specific input vector. | |
| --- | --- | --- | --- | --- |
| GLM | | Parametric | A specific theory or model is fitted to the data | Pomeranz et al. 2019 |

# Methods

## Machine learning models for predicting species interactions from trait-matching

Throughout this paper, we consider that empirical observations of species interactions may be available either as binary (presence-absence) or weighted (counts, intensity, interaction frequencies) data. The objective for the models is to predict those plant-pollinator interactions based on the species' traits. We selected seven classes of ML models, either because they were previously used for trait-matching, or because the general ML literature suggest that they should perform well for this task (Table 1). For more details on the respective models, see the column "Design principle" and the cited literature in Table 1, and the Supporting Information S1 in the Appendix.

Each of the ML models in Table 1 includes model-specific tuning parameters (so-called hyperparameters, for instance to control the model's learning behavior) that can be adjusted by hand or optimized. To factor out idiosyncrasies due to the choice of these parameters, we optimized each models' hyperparameters with a random search in 30 (20 for empirical data) steps (see also Bergstra & Bengio 2012), with nested cross-validations to avoid overfitting (for details see Appendix S1). Furthermore, ML models often perform poorly with imbalanced classes (proportion of plant-pollinator interactions to no plant-pollinator interactions is extremely low/high, Krawczyk 2016). To address this, we applied the standard approach of oversampling observed plant-pollinator interactions when their proportion (compared to plant-pollinator pairs without an interaction) was lower than 20%.

To compare ML with traditional regression models, we fitted GLMs (binomial GLM for presence-absence plant-pollinator interactions and Poisson GLM for plant-pollinator interaction counts), using all traits and all their possible two-way interactions as predictors (and plant-pollinator interactions as response). Analyses were conducted with the statistical software R (R Core Team, 2019). The R package mlr (Bischl et al. 2016, version 2.12) was used for hyperparameter tuning and cross-validation of our ML models.

## Simulating plant-pollinator interactions

To assess predictive and inferential performance of the models, we created a minimal simulation model for plant-pollinator interactions. The model assumes that the interaction probability between individuals of plants (group A) and pollinators (group B) arises from a Gaussian niche, matching the logarithmic ratio of the plant and pollinator traits (e.g. log ($A_i/B_j$), centered around 1). The logarithmic ratio ensures a symmetrically shaped interaction niche, see Fig. S1. The niche value is multiplied by a weight to allow modifying the interaction strength independent of the niche width, and thus to control the overall trait-matching effect signal. Plant and pollinator abundances can either be drawn from an exponential distribution or a uniform distribution, to examine the effects of uneven abundance distributions and rare species. The expected number of observed interactions (i.e. their probability, $P_{interaction}$) was then calculated as the interaction probability times the interaction partner's abundances times the observation time. Observation times were adjusted to standardize the proportion of plant-pollinator interactions to no plant-pollinator interactions. To create the final interaction counts, we sampled from a Poisson distribution with $\lambda = P_{interaction}$. For presence-absence species interactions (1 = interaction, 0 = no interaction), we set all counts > 0 to 1.

Our default simulation scenario used 50*100 (plants*pollinators) for the simulated plant-pollinator networks. To remove obstacles such as class imbalance, we adjusted the observation duration to have a class proportion of ≈ 40% for plant-pollinator interactions to no plant-pollinator interactions. The absence of interactions cannot be observed explicitly, and we speculate that most empirical datasets consist of observed species interactions (and possible non-interactions are inferred afterwards), thus we removed species with no observed plant-pollinator interaction.

## Comparison of predictive performance

### Predicting species interactions in simulated plant-pollinator networks

To assess predictive performance, we simulated reference data with six traits for each plant and pollinator. A possible issue with measuring predictive performance is that hidden correlations or structure in the data can lead to seemingly higher-than-random predictive performance even on random data (e.g. Roberts et al. 2017). To check that this is not the case, we created a first baseline scenario, consisting of equal species abundances and no trait-trait interactions (no trait-matching, the latter was achieved by setting the trait-trait interaction niche extremely wide). A second issue is that interactions of rare species will be less frequent than those of abundant species. As a result, models can achieve higher-than-random performance even without any trait-trait interactions when species abundances are uneven. To ensure that the performance of our models exceeds these trivial performance levels, we created a second baseline scenario with exponential abundance distributions, but without trait-matching.

For the trait-matching scenario, we simulated networks with even abundance distributions and three trait-trait interactions (A1-B1, A2-B2, and A3-B3), each with a weight of ten. The scale parameter controlling the niche width was randomly sampled between 0.5 and 1.2 for simulating varying degrees of specialization in ecological networks (cf. Blüthgen et al.

2007). The even abundance distributions assumed here are unrealistic to some extent, but allow a better contrast between the models (because abundance effects are removed). In the case studies, we consider real abundance distributions. Other than that, the trait-matching scenario used the same parameter settings as the baseline scenarios (network size 50*100, ≈ 40% class balance). To test additionally for the effect of network sizes and observation time, we also varied network size to 25*50 and 100*200 (plants*pollinators) setting and proportions plant-pollinators interactions to ≈10 %, ≈ 25 %, and ≈ 40 % one-factor-at-a-time from the base setting.

## Case study 1 - Predicting plant-pollinator interactions

Our first case study uses data from a global database of crop-pollinator interactions, assembled from 1607 published studies from 77 countries worldwide (details see section Data Accessibility). Of these, we selected only crops that appeared at least two times at different geographical locations, resulting in 80 crops with 256 entries for pollinators.

The database lists five pollinator traits: guild (bumblebees, butterflies etc.), tongue length, body size, sociality (yes or no), and feeding behavior (oligolectic, polylectic, or parasitic). In case of sexual dimorphism, the female measures were taken. Plants are described by ten traits: type of plant (arboreous or herbaceous), flowering season, flower diameter, corolla shape (open, campanulate, or tubular), flower color, nectar (yes or no), bloom system (type of pollination: insects, insects/wind, or insects/birds), self-pollination (yes or no), inflorescence (yes, solitary, solitary/pairs, solitary/clusters), and composite flowers (yes or no). Flower diameter, body size, and tongue length were provided as continuous traits (see Tables S1 and S2 for detailed information). When traits for a species were available from different sources, they were averaged. We filled missing trait values with a multiple imputation algorithm based on random forest (Stekhoven & Bühlmann 2012). We used all available traits as predictors in our models.

## Measure predictive performance

To assess the models' predictive performance on the simulated plant-pollinator networks, we used the area under the receiver operating characteristic curve (AUC, measures how well the models are able to distinguish between plant-pollinator interaction and no plant-pollinator interaction regardless of classification threshold) and true skill statistic (TSS, which assess the predictive performance under a specific classification threshold, see Allouche, Tsoar, & Kadmon, 2006) for presence-absence, and spearman's correlation for interaction frequencies. Because the TSS for the empirical plant-pollinator database (case study 1) were similar, we additionally calculated classification threshold-dependent performance measurements: accuracy (proportion of correct predicted labels), sensitivity (recall), precision, and specificity (true negative rate). Classification thresholds were optimized with TSS. The interpretation of these statistics is as follows: if our focus is to detect plant-pollinator interactions, we want to achieve a high true positive rate (sensitivity) with an acceptable rate of false positives in the as true predicted labels (precision). Specificity estimates the rate of true negatives of all predicted negatives (no plant-pollinator interaction).

## Measuring accuracy for inferring causal traits

### H-statistics for inferring causal traits

We used the H-statistic (Friedman & Popescu 2008) to infer causally responsible trait-trait interactions from the fitted ML models. The idea of this algorithm is similar to the principle of partial dependence plots. The H-statistic estimates the variance of the model's response caused by two traits separately (main effects) compared to the variance caused by the two traits combined partial function (trait-trait interaction). The H-statistic is scaled to [0,1]. A high value indicates that the interaction is the main reason for the variance in the response (probability for plant-pollinator interactions and counts for plant-pollinator interaction counts).

## Inferential performance in simulated plant-pollinator networks

To assess the accuracy with which causal trait combinations can be identified from the fitted models via Friedman's H-statistic, we considered 25*50 (plants*pollinators) species networks with one, two, three and four trait-trait interactions (always six traits for each group but varying number of trait-trait interactions that correspond to trait-matching), and equal interaction strength. We replicated the simulated plant-pollinator networks eight to ten times. The reason for choosing a smaller network size than for the predictive analysis was the computational cost of the H-statistics, which made applying a large number of replicates to larger networks computationally prohibitive.

The resulting networks had a 'real' observed size of 800 - 1200 data points (we removed two networks with four true trait-trait interactions, because they had under 20 remaining samples after removing species with no plant-pollinator interactions at all). We fitted RF, BRT, DNN and kNN (the top predictive models) on the 76 simulated networks, 38 for presence-absence plant-pollinator interactions and 38 for plant-pollinator interaction counts (with uniform species abundances). For each sample, we calculated the H-statistic for all possible trait-trait interactions between the two species' groups. We calculated for each, the averaged true positive rate (true trait-trait interaction in found interactions with highest H-statistic) over the eight/ten repetitions. In a second step, based on our interim results (see results) we repeated the procedure with BRT and DNN for 50 * 100 (plants*pollinators) simulated networks (see Appendix S1 for details regarding model fitting).

For GLMs, we selected the n (n = number of true trait-trait interactions) predictors with lowest p-value to calculate the true positive rate.

## Case study 2 – Inferring trait-matching in a plant-hummingbird network

As a case-study for inferring causally responsible traits, we used a dataset of plant-hummingbird interactions from Costa Rica. Plant-hummingbird networks are characterized by particularly strong signals of trait-matching (Vizentin-Bugoni, Maruyama, & Sazima, 2014). Maglianesi et al. (2014) filmed and analyzed plant-hummingbird interactions at three elevations in Costa Rica (700 hours of observations on 50 m.a.s.l; 695 hours of observation on 1000 m.a.s.l; 727 hours of observations on 2000 m.a.s.l). The resulting network consisted of 21*8, 24*8, and 20*9 plant and hummingbird species, respectively. To predict plant-pollinator interactions, we used bill length, bill curvature, body mass, wing length, and tail length of hummingbirds, and corolla length, corolla curvature, inner corolla diameter width, and external corolla diameter width of plants. Flower volume was calculated by corolla length and external diameter (Maglianesi et al. 2014). We used all available traits because the ML models should automatically learn trait-trait interactions.

We fitted the BRT with a Poisson maximum likelihood estimator and RF with a root mean squared error (RMSE) objective function (we did not log count data). We optimized DNNs with Poisson and negative binomial likelihood loss functions. We trained models on each elevation and on combined elevations (e.g. Low, Mid, High, Low-Mid-High, for details see Appendix S1). We calculated for the Low, Mid, High and Low-Mid-High models interaction strengths (H-statistics) for all possible trait-trait interactions (with trait-trait interactions within hummingbird/plant group). We checked the eight trait-trait interactions with highest interaction strengths for their biological plausibility by reviewing relevant literature.

# Results

## Predictive performance

### Predictive performance in simulated plant-pollinator networks

In the first baseline scenario (no trait-matching and equal species abundances), all models performed as expected for random plant-pollinator interactions, with AUC ≈ 0.5, TSS ≈ 0, and Spearman Rho factor ≈ 0 for both for presence-absence data and count data (Fig. 2), indicating that our cross-validation setup is accurate. In the second baseline scenario (no trait-matching and networks with uneven species abundances), models achieved a TSS between 0.0-0.38, AUC between 0.64-0.76, and Spearman Rho factor of between 0.26-0.5 (Fig. 2). The latter provides an indication, also with respect to existing literature, of what performance values can be achieved through imbalance of the data alone, even if there is no trait-matching.

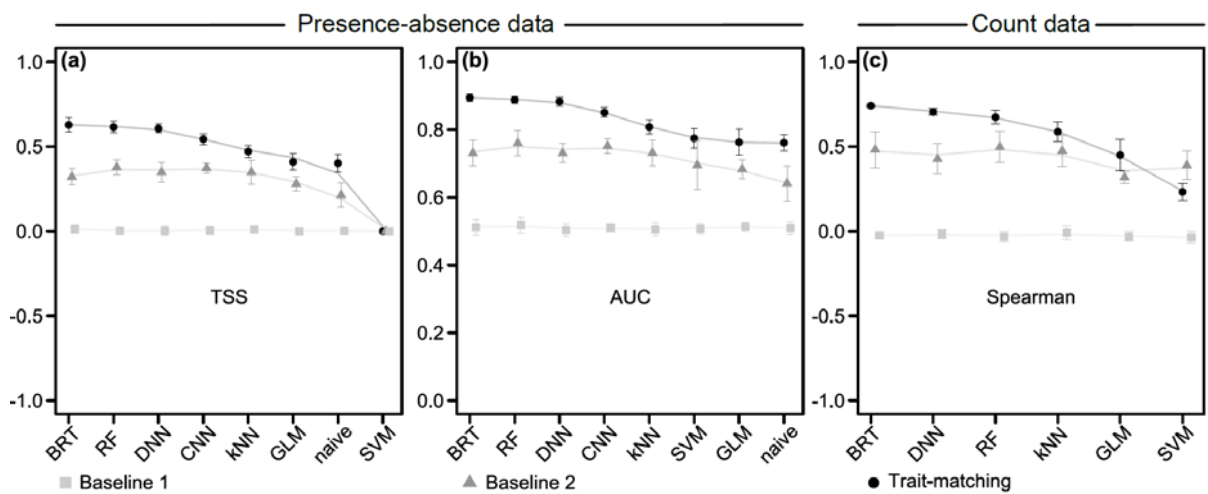

**Figure 2:** Predictive performance of kNN, CNN, DNN, RF, BRT, naive Bayes, GLM and SVM with simulated plant-pollinator networks (50 plants * 100 pollinators) for baseline scenarios with random interactions and even (baseline 1, squares) or uneven species abundances (baseline2, triangles, respectively), and trait-based interactions with even species abundances (circles). Predictive performance was measured by TSS (a) and AUC (b) for binary interaction data; and Spearman Rho factor (c) for interaction counts. Lowest predictive performance corresponds to zero for TSS, AUC, and Spearman Rho factor.

For simulated networks with strong trait-matching and even abundances, all ML models except SVMs achieved higher TSS, AUC, and Spearman Rho than for the baseline scenarios (Fig. 2). Moreover, DNN, RF, and BRT achieved a higher TSS (0.61-0.63) than GLMs (0.41). SVM, naïve Bayes, kNN were around GLM's performance or lower (Fig. 2).

While all models improved their predictive performance with increasing network sizes with count data (Fig. S2c), only DNN, RF, and BRT improved their performance with increasing network sizes with presence-absence plant-pollinator interactions (Fig. S2 a-b). Prolonging the observation time (i.e. creating more plant-pollinator interactions and thus reducing data imbalance) generally increased the models' performances (Fig. S2d-f).

## Predicting species interactions in a global crop-pollination database

After fitting the models to real data from a global crop-pollination database, we calculated AUC, TSS and additional performance measures (Fig. 3, Table S4) on the left-out samples. kNN achieved the highest TSS (0.36), RF achieved the highest AUC (0.73), and naïve Bayes achieved highest TPR, followed by CNN. Overall, RF achieved the overall best predictive performance with highest AUC and second highest TSS (Fig. 3, Table S4).

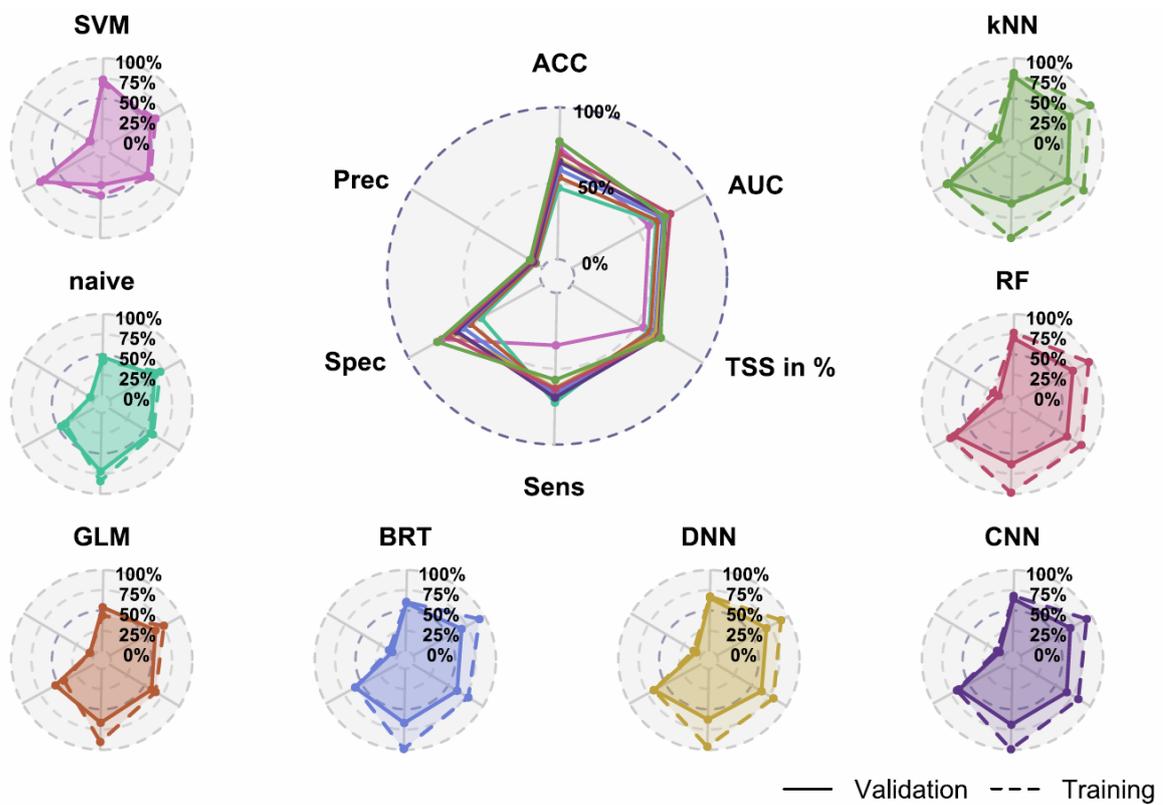

**Figure 3:** Predictive performance of different ML methods (naive Bayes, SVM, BRT, kNN, DNN, CNN, RF) and GLM in a global database of plant-pollinator interactions. Dotted lines depict training and solid lines validation performances. Models were sorted from left to right with increasing true skill statistic. The central figure compares directly the models' performances. Sen = Sensitivity (recall, true positive rate); Spec = Specificity (true negative rate); Prec = Precision; Acc = Accuracy; AUC = Area under the receiver operating characteristic curve (AUC); TSS in % = True skill statistic rescaled to 0 – 1.

# Inference of causal trait-trait interactions

## Inference of causal trait-trait interactions in simulated networks

In the second analysis step, we tested the ability of the H-statistics to infer the trait-trait interactions causally responsible for plant-pollinator interactions from the fitted models. In simulated networks, RF and BRT achieved highest true positive rates (Fig. 4). For presence-absence plant-pollinator interactions, RF, DNN and BRT exceeded GLM performance with an averaged true positive rate of 70% to 80% over one to four true trait-trait matches (Fig. 4a, the models were able to identify most of the true trait-trait interactions). For plant-pollinator interaction count data, only RF achieved a higher true positive rate than GLM (Fig. 4b). However, it should be noted that the good GLM performance hinged on simulations with 1-3 trait-trait interactions and decreased most strongly of all algorithms with the number of trait-trait interactions (Fig. 4).

When increasing network size (from 25*50 to 50*100), DNN and BRT improved their overall performance to 70 – 95 % and 87 – 98 % for presence-absence networks (Fig. S3a), but showed a lower TPR for count data (Fig. S3b).

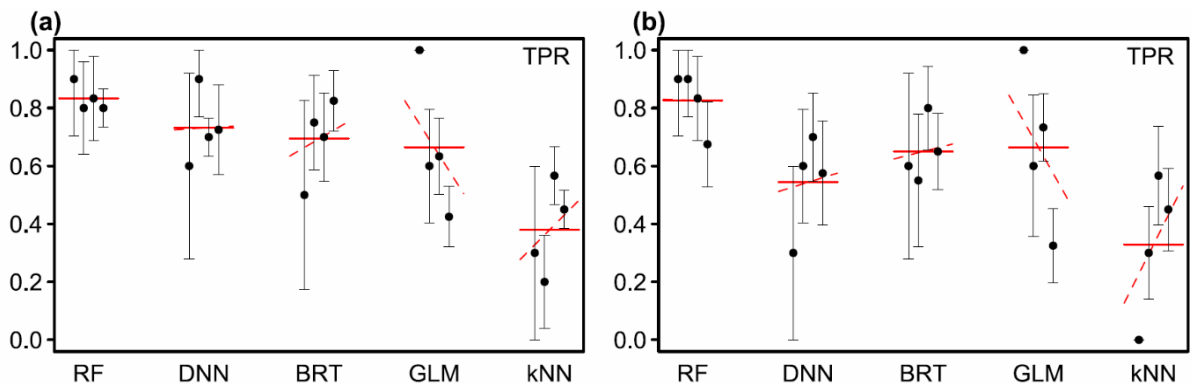

**Figure 4:** Comparison of the top predictive models' (RF, DNN, BRT, kNN, and GLM) abilities to infer the causal trait-trait interaction structure in simulated networks, using presence-absence data (a) and count data (b). The four values associated with each algorithm represent the mean true positive rate (TPR, dot) and its standard error (error bar) for the four interaction scenarios (one to four true trait-trait interactions in the simulations). The values were calculated based on 8-10 replicate simulations each. Solid red lines display the mean TPR across all four scenarios, dotted red lines show a linear regression estimate of TPR against the number of true trait-trait interactions.

# Inference of causal trait-trait interactions in a plant-hummingbird network

In a second case study, we computed interaction strength (H-statistic) for all possible trait-trait interactions in plant-hummingbird networks (Fig. S5). The seven trait-trait interactions with highest interaction strength were identified by RF (Fig. 5b). These interactions also achieved highest predictive performance (Fig. S4). The four trait-trait interactions with highest interaction strength identified by BRT were in accordance with the ones that RF identified (Fig. S5).

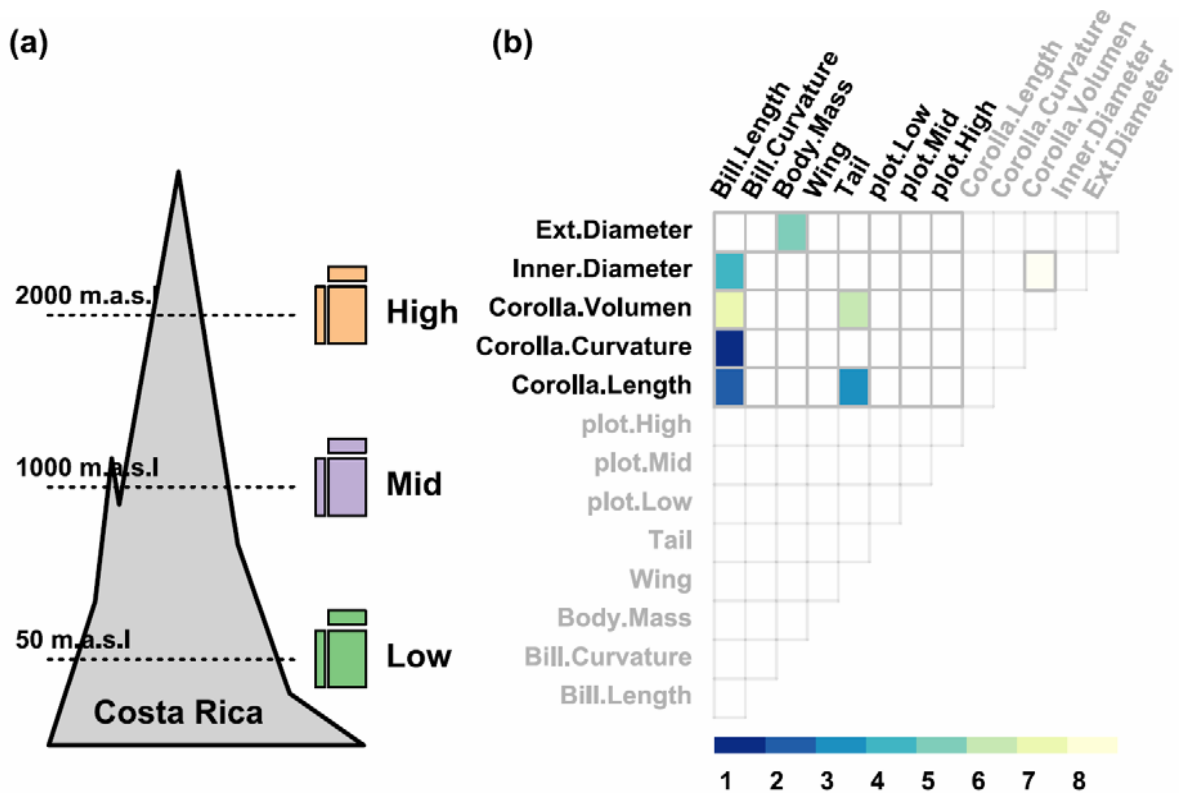

**Figure 5:** a) Elevation profile for the three plant-hummingbird networks in Costa Rica (details see Maglianesi et al. 2014; b) The eight strongest trait-trait interactions (blue – yellow gradient) inferred with the H-statistic from RF models fitted to the combined plant-hummingbird network (colors code the ranking of strengths). Corolla length – bill length and corolla curvature – bill length had the highest interaction strength.

RF and BRT identified corolla length-bill length, corolla curvature-bill length, inner diameter-bill length, and external diameter-body mass as the most important trait-trait interactions (Fig. 5b, Fig. S5). The models identified varying trait-trait interactions for networks at different elevations, but corolla and bill associations tended to be most important across elevations (Fig. S5).

# Discussion

We assessed the ability of seven ML models, plus GLM as a reference, to predict plant-pollinator interactions based on their traits. In a second step, we tested whether it is possible to identify the causally responsible trait-trait interaction structure (trait-matching) from the fitted models. Our main results are that the best ML models (RF, BRT, and DNN) outperform GLMs to a substantial degree in predicting plant-pollinator interactions from their traits, and that it is possible to identify the trait-trait interactions causally responsible for plant-pollinator interactions from the fitted models with satisfying accuracy. The best ML models outperformed the simpler GLMs particularly for more complex trait-trait interaction structures, for which GLM performance dropped sharply.

## Comparison of performance in predicting species interactions

In our analysis of predictive performance, we found that ML models such as RF, BRT, and DNNs exceeded GLM performance for predicting plant-pollinator interactions from trait-matching data. They also worked surprisingly well with small network sizes (25*50, Fig. S2a), such that performance did not increase substantially for larger networks (50*100, 100*200, Fig. S2 a-c).

An important point, also for comparing our performance indicators to the literature, is that all algorithms can achieve higher than naïve random performance (e.g. AUC of 0.5) when species distributions are uneven, even when plant-pollinator interactions are not tied to traits (Fig. 2). These results, in line with earlier findings (e.g. Aderhold et al. 2012; Canard et al. 2014), highlight the importance of considering abundance when analyzing network structures: frequent species tend to have more observed interactions, and this effect might interfere with the trait-matching signal (e.g. Olito & Fox 2015). While the trait-matching effect may influence which plant-pollinator interaction is feasible, the species abundance effect determine the actual observed plant-pollinator interactions. Without adjusting observed plant-pollinator interactions for species abundances, it is difficult to separate the contributions of abundance and trait-matching to predictive performance (Olito & Fox 2015).

Observation time and type are further critical factors in ecological network analysis. Short observation times often lead to sparse networks with many unobserved plant-pollinator interactions, potentially creating biases in the analysis. Moreover, few plant-pollinator interactions result in data with imbalanced class distributions, presenting challenges for many ML methods (Krawczyk 2016), which is also reflected in our results (Fig. S2d-f). On the other hand, too long observation times could also negatively affect predictive performance, in particular when using binary links. The reason is that, given sufficient time, even weak links will be included in the network, potentially reducing the models' ability to identify the essential traits. Count data is more robust to these problems, and as our approach is equally applicable with count data, this data type seems generally preferable (see also Dormann & Strauss 2014).

While the ML models detected the important trait-trait interactions automatically, GLMs were pre-specified with all possible two-way trait-trait interactions. To check that the resulting high complexity did not disadvantage them unduly, we additionally confirmed that AIC selection on their interaction structure did not increase their predictive performance. We therefore believe that their lower performance is either explained by the fact that GLMs are not flexible enough to capture the complex form of the trait-matching structures (see also Mayfield & Stouffer 2017), or that ML methods are more successful than AIC variable selection in addressing overfitting induced by the high combinatorial number of possible trait-trait interactions. These results mirror findings in the literature: while a few studies showed that GLM can predict species interactions based on trait-matching (e. g. Gravel et al. 2013), most studies struggled in predicting species interactions with the trait-matching signal alone (e.g. Pearse & Altermatt 2013; Brousseau, Gravel, & Handa, 2018; Pomeranz et al. 2019). We speculate based on our results that previous studies based on GLMs may have underestimated the importance of trait-matching considerably, unless a very small number of trait-trait interactions (1-2) is dominantly responsible for the structure of the networks.

Previous studies often showed improved performance in predicting species interactions by using phylogenetic predictors, serving as proxies for unobserved traits (see Morales-Castilla et al. 2015). The drawback, however, is that such phylogenetic proxies can be hard to interpret in the context of specific ecological hypothesis of why species interact (see Díaz et al. 2013). For example, a phylogenetic signal could arise both as a result of trait-matching (because traits tend to be phylogenetically conserved), or as a result other genetically coded preferences for particular interactions that are not accessible as traits. Based on our results, we expect that the relative importance of phylogenetic proxies will decrease when using appropriate ML models, which could help to better explore to what extent species interactions are determined by measurable functional traits.

We found that the models' predictive performance was lower for the empirical plant-pollinator database than for the simulated networks. There are several plausible reasons for this. Firstly, trait-matching rules may change over scales (Poisot, Stouffer, & Gravel, 2015). As the database consists of globally observed plant-pollinator interactions, this may complicate the identification of a common trait-matching signal. Secondly, the high share of discrete predictors and high-class imbalance is likely to negatively affect the predictive performance. Despite these obstacles, kNN, RF, and CNN achieved > 0.3 TSS, and CNN and RF > 70 % AUC (Fig. 2, Table S4), much higher than null expectation, and consistent with results from the simulated networks. While it may be possible to improve GLM performance by manual selection of predictors, we also find that the case study highlights that algorithms such as RF and BRT are more parsimonious and robust in their use than a GLM which further suffered convergence problems.

# Causal inference of trait-matching

To infer trait-trait interactions causally responsible for species interactions, we used the H-statistics. We found that this method, coupled with RF, DNN, and BRT, could identify around 90% of the true trait-trait interactions in simulated plant-pollinator networks (Fig. 4, Fig S5). Increasing the network size improved the detection accuracy of true trait-matches for BRT and DNN (Fig. S3). When increasing the number of trait-trait interactions, the approach outperformed GLMs (Fig. 2).

Our results demonstrate that identifying trait-matching from fitted models with the H-statistic works, but it also comes with drawbacks. The H-statistic depends on partial dependencies (Friedman & Popescu 2008) and is therefore sensitive to collinearity (see Apley 2016). Other alternative approaches (e.g. Apley 2016) might overcome this limitation. Moreover, the H-statistic is extremely computationally expensive, which is the reason why we tested it only on small network sizes (25*50 species). Neither of these issues, however, would change the balance in favor of GLMs, which are prone to collinearity issues, too. To make sure that GLMs are not unjustly disfavored, we additional tested if AIC selection or choosing causal traits based on regression estimates instead of p-values would change the results, but neither improved inferential performance. In summary, we think that ML models are the better choice, not only for predictions, but also for causal inference in this setting. Future research should, however, focus on testing and advancing methods for the causal analysis of fitted models.

Analyzing plant-hummingbird networks with RF, we highlighted the seven trait-trait interactions with highest interaction strength (Fig. 5b). The inferred trait-trait interactions are highly plausible for the following reasons: i) RF showed high accuracy with low consistent errors in the simulated networks (Fig. 4). ii) The identified trait-trait interactions are ecologically plausible (Fig. 5b): Trait matches with highest interaction strength (corolla length-bill length and corolla curvature-bill length) are in line with previous findings that emphasize their importance in plant-hummingbird networks (Temeles et al. 2009;

Maglianesi et al. 2014; Vizentin-Bugoni, Maruyama, & Sazima, 2014; Weinstein & Graham 2017). Collinearity of traits likely explains other matches. For instance, body mass is positively correlated with tail length, explaining why corolla volume was associated with tail length. These results further support the view that it is possible to infer trait-matching with ML in ecologically realistic settings, without a priori assumptions.

Estimated trait-trait interactions in the plant-hummingbird networks differed for the three elevations, but the match of corolla length–bill length was generally most important (Fig. S5). Maglianesi et al. (2014) and (2015) reported similarly varying trait-trait interactions in plant-hummingbird networks across elevations, consistent with our results. While interactions in ecological networks vary over scales (Poisot, Stouffer, & Gravel, 2015), a common backbone is assumed (Mora et al. 2018). With corolla length - bill length, identified by RF and BRT with highest interaction strength (Fig. 5b, Fig. S5), we speculate that we identified with ML the central trait-matching phenomenon in plant-hummingbird networks.

# Conclusions

In conclusion, our study demonstrates that RF, BRT, and DNN exceeded GLM performance in predicting plant-pollinator interactions from trait information. ML models could also identify causally responsible trait-trait interactions with a higher accuracy than GLMs. The ability to automatically extract species interactions from observed networks and traits, and causally interpreting the underlying trait-trait interactions, makes our approach, which we provide in an R package, a powerful new tool for ecologists.

While we considered only plant-pollinator networks in this study, our method could be applied to other types of species interaction networks such as any mutualistic and antagonistic interactions in complex food webs (this is also supported by Desjardings-Proulx et al. 2017). In either of these ecological network types, there are ample opportunities for further analyses, for example how species interactions will change under global change or how species interactions will rewire in novel communities with reshuffled species and trait composition (Bailes et al. 2015; see Kissling & Schleuning 2018). By identifying crucial rules of trait-matching between species, our approach can give insights into how biotic interactions shape community assembly and also contributes to the identification of Essential Biodiversity Variables in the context of global change (Kissling et al. 2018).

# Acknowledgements

Maria A. Maglianesi recorded interaction and trait data of plants and hummingbirds in Costa Rica. We would like to thank Johannes Oberpriller and Lukas Heiland, as well as two anonymous reviewers for their valuable comments and suggestions. VB acknowledges funding for the assembly of the global plant-pollinator database (case study 1) by Bayer CropScience.

# Authors' contributions

MP and FH conceived the ideas and designed methodology; VB, AMK and MS provided data; MP performed the analyses. MP and FH wrote the first draft of the manuscript. All authors contributed critically to the completion and revision of the manuscript.

# Data Accessibility

The plant-hummingbird data associated with this study is available at https://doi.org/10.6084/m9.figshare.3560895.v1. The global plant-pollinator database used with this study is available at https://doi.org/10.6084/m9.figshare.9980471.v1. The analysis and the Trait-matching R package is available at http://doi.org/10.5281/zenodo.3522854 (https://github.com/TheoreticalEcology/Pichler-et-al-2019).

# Appendix

## Plant-pollinator simulation

In our simulated plant-pollinator the trait-trait interactions (trait-matching) result from a Gaussian niche. To ensure that fraction of two traits (e.g. A1 = 0.5 for species i and B2 = 0.7 for species j) has the same distance to zero from both sides of the mean, we used the logarithmic fraction: log(0.5/0.7) ≈ 0.34 and log(0.7/0.5) ≈ -0.34. The maximum for this fraction is A1 equal B1: log(0.5/0.5) = 0 and obtains the highest interaction probability of the Gaussian niche (μ = 0).

The difference between logarithmic and no logarithmic is shown in Fig. S1. For no logarithmic faction, equal absolute distances for the optimal fraction A1 equal B1 cannot be guaranteed.

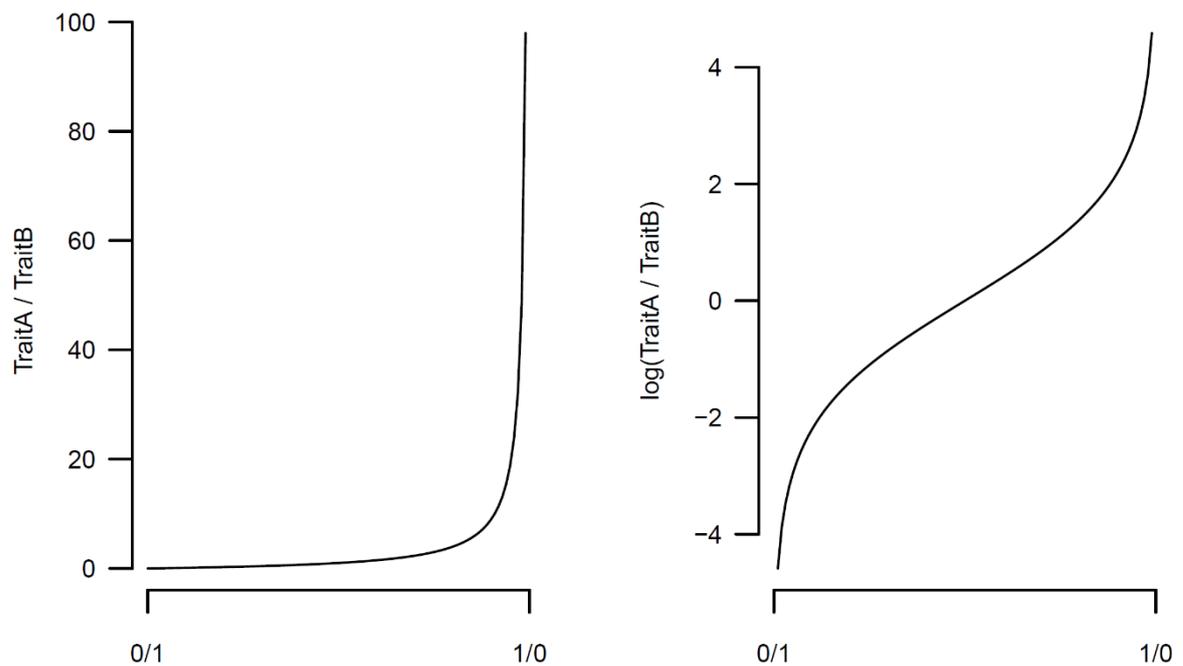

**Figure S1:** Unlogged and logged fraction of two trait-trait matches.

# Plant-pollinator databases

The global plant-pollinator database is available on:

https://doi.org/10.6084/m9.figshare.9980471.v1

**Table S1:** Detailed information on plant traits for the plant-pollinator database

| Trait | Type | Levels | Additional information |
|---|---|---|---|
| Type | Discrete | arboreous, herbaceous | |
| Flower season | Discrete | sprisum, summer, spriaut, spring, autspri, sumspri, autumn, year, sumaut, wispring, winter | Describes the seasonal range. For instance, sprisum correspond to spring – summer range. |
| Flower diameter | Continuous [mm] | | |
| Flower corolla | Discrete | campanulate open, tubular | |
| Nectar | Discrete | Yes, No | Whether flower contains nectar or not. |
| Flower color | Discrete | white, yellow, purple, pink, green, blue, red | |
| Bloom system | Discrete | insects, insects/bats, insects/bats, insects/birds | Type of pollinator |
| Self-pollination | Discrete | Yes, No | |
| Inflorescence | Discrete | solitary, solitary/clusters, solitary/pairs, yes | |
| Composite | Discrete | Yes, No | |

**Table S2:** Detailed information on pollinator traits for the plant-pollinator database

| Trait | Type | levels |
|---|---|---|
| Guild | Discrete | andrenidae, bumblebees, butterflies, coleoptera, cuckoo bees, flies, honey bees, moths, other, other bees, stingless bees, sweat bees, syrphids, wasps |
| Tongue | Continuous [mm] | |
| Body | Continuous [mm] | |
| Sociality | Discrete | Yes, No |
| Feeding | Discrete | oligolectic, parasitic, polylectic |

# Model Training

## Comparison of predictive performance under different network characteristics

We standardized the features before fitting. We fitted models' hyper-parameter in 30 random tuning steps. We used nested cross-validation, five times outer and three times inner, to maximize generalization and to estimate overfitting. We applied cross-validation by putting an insect with all its possible plant interaction partners out. We regularized deep neural network (DNN) and convolutional neural network (CNN) with dropout (rate = 0.2) for presence-absence plant-pollinator interactions and for plant-pollinator interaction counts with batch normalization (interim results showed that batch normalization worked better for plant-pollinator interaction counts).

**Table S3:** Overview of hyper-parameters we tuned in the predictive performance comparison.

| Model | Hyper-parameter | Range | R package |
| --- | --- | --- | --- |
| RF | mtry | 2 - (number of features-1) | randomForest (Liaw and Wiener, 2002), ranger Wright and Ziegler, 2017) |
|  | nodesize | 2 - 50 |  |
|  | replace | Yes/No |  |
| DNN | learning rate | 0.1 - 0.0001 | keras (Chollet *et al.* 2017), tensorflow (Abadi *et al.* 2015) |
|  | hidden nodes | 5 - 50 |  |
|  | number of layers | 1 - 5 |  |
|  | bias | Yes/No |  |
|  | optimizer | Sgd, adam, rmsprop |  |
|  | decay (optimizer decay) | 0.9 - 0.99 |  |
|  | number of layers | 1 - 6 |  |
| CNN | learning rate | 0.1 - 0.0001 | keras (Chollet *et al.* 2017), tensorflow (Abadi *et al.* 2015) |
|  | hidden nodes in fc layer | 10 - 80 |  |
|  | number of kernels (filter) | 8 - 30 |  |
|  | pooling | max/average |  |
|  | decay (optimizer decay) | 0.9 - 0.99 |  |
| BRT | booster | gbtree, dart | Xgboost (Chen and Guestrin 2016) |
|  | sample type | uniform/weighted |  |
|  | normalize type | tree/forest |  |
|  | eta | 0.01 - 0.5 |  |
|  | max depth | 1 - 10 |  |
|  | lambda | 0.1 - 10 |  |
|  | alpha | $2^{-10} - 2^{5}$ |  |
|  | min child weight | 0 - 10 |  |
|  | number of rounds | 1 - 500 |  |
|  | dropout rate | 0 - 0.2 |  |
|  | skip dropout | 0 - 0.3 |  |
| kNN | k | 1 - 10 | kknn (Schliep and Hechenbichler 2014) |
|  | kernel | Rectangular, triangular, epanechnikov, optimal |  |
| naive Bayes | laplace | 0 - 6 | e1071 (Meyer *et al.* 2019) |
| SVM | lambda | 0.01 - 20 | liquidSVM (Steinwart and Thomann 2017) |
|  | gamma | 0.01 - 20 |  |
|  | kernel | rbf, poisson |  |

We used early stopping (patience = 10) and a callback to reduce loss on plateaus to optimize training in DNNs.

The study was done with the statistical computing software R (R Core Team 2019). Tuning and cross-validation was implemented in our Trait-Matching package with the help of the R package mlr (Bischl et al. 2016, version 2.12).

## Description of the ML algorithms

*Random Forest (RF) and boosted regression trees (BRT)*

RF and BRT are based on classification and regression trees. During training, the dataset is split by specific values of the predictors' distributions. In each split, the predictors are searched for the value that split the predictors and response so that the response's variance (regression) is minimized or the accuracy is maximized (classification).

BRT fit hundreds of trees sequentially on the data. During training, the first tree has the observed labels as response while the subsequent trees have the residual errors as responses. This is known as gradient boosting.

The RF algorithm compromise two random steps: (1) RF fits hundreds of trees on bootstrap samples and (2) in contrast to BRT or trees, the RF algorithm randomly subsamples the predictors in each split (RF has to choose the best split value from a subset of predictors). Predictions in RF and BRT are averaged predictions (regression) or classification labels by majority voting.

*DNN*

Deep neural networks are based on artificial neural networks. The input predictors are mapped over fully connected layers, each input node is connected to all nodes in the layer, to output nodes whose numbers correspond to the number of response classes (classification) or to one output node (regression). DNNs can consist of hundreds of hidden layers in which each node in a layer is connected to all nodes in the following layers. During training, the weights are updated by backpropagation: The weights in each layer are updated in dependence of the error in the output node by gradient descendance (More detailed: A loss function specifies the error, e.g. entropy (classification), mean squared error (regression)).

*CNN*

Convolutional neural networks are based on convolutional layers. Compared to DNNs, CNN can handle topological inputs such as 1-D sequences or 2-D images. A convolutional layer consists of kernels, usually small n*n weight matrices, that apply the 'convolution' function over the input space, i.e. they run over the input space with a specific step size and use actually cross-correlation. The outputs of these kernels are feature maps. The topological information is still conserved within the feature maps. After a pooling layer (max or average reduction of feature maps with windows (e.g. 2*2) and a specific step size, additional convolutional layers can follow. After the last convolutional layer, the features maps are collapsed and connected to a fully connected layer, followed by an output layer. Training is congruent to DNNs.

*k-nearest-neighbor (kNN)*

The K-nearest-neighbor algorithm computes the distance between all observations in the feature space. A new observation is classified by majority vote of the k-nearest neighbors. For regression, the predicted response is the averaged response of the k-nearest neighbors. If the response is non-linear, the feature space can be transformed with a kernel (e.g. gaussian kernel).

*naïve Bayes*

naive Bayes is based on the bayes theorem. It learns the probability of data point $X_i$ belonging to class $y_i$ given the input vector $X_i$ (features/predictors). Naïve bayes is used only for classification.

*Support vector machines (SVM)*

Support vector machines optimize hyperplanes to separate the observations into their response classes (classification) or the averaged responses (regression) in the feature space. The term 'support vector' refer to the computational benefit that only observations close to the plane are used to optimize the plane. To overcome the issue that only linear separable tasks work well with SVMs, the feature space can be transformed with a kernel (the kernel trick, e.g. gaussian kernel).

## Fitting models for inferring responsible trait-trait interactions

We optimized networks' observation time in such manner that the class proportion of presence interactions was around 40 %. We filtered networks for the condition that at least each insect should have one observed interaction. Only networks with a sample size with 50 % of the original size were used. We fitted random forest (RF), boosted regression tree (BRT), DNN and k-nearest-neighbor (kNN) in 50 random tuning steps (Table S1). We used holdout validation (split 80:20) for outer and inner validation. We applied the same procedure for the 50*100 networks with DNN and BRT.

For inferring responsible traits, we used a grid size equal to the maximum number of rows in the data in the 25 * 50 (plants * pollinators) networks. For the 50 * 100 networks, we used a grid size of 500.

## Additional Results

## Comparison of predictive performance

**Table S4:** Results from the comparison of machine learning models for their predictive performance in an empirical plant-pollinator network. Rows are sorted according to TSS.

AUC = area under the receiver operating characteristic curve; acc = Accuracy; tss = true skill statistic.

| auc | f1 | bac | acc | fdr | precision | specificity | sensitivity | tss | method |
|---|---|---|---|---|---|---|---|---|---|
| 0.577 | 0.09 | 0.548 | 0.736 | 0.948 | 0.052 | 0.75 | 0.346 | 0.096 | SVM |
| 0.628 | 0.096 | 0.59 | 0.47 | 0.948 | 0.052 | 0.458 | 0.723 | 0.181 | naive |
| 0.643 | 0.104 | 0.602 | 0.541 | 0.942 | 0.058 | 0.538 | 0.666 | 0.204 | GLM |
| 0.679 | 0.131 | 0.631 | 0.592 | 0.925 | 0.075 | 0.592 | 0.67 | 0.261 | BRT |
| 0.696 | 0.124 | 0.642 | 0.656 | 0.93 | 0.07 | 0.658 | 0.627 | 0.285 | DNN |
| 0.701 | 0.128 | 0.665 | 0.641 | 0.929 | 0.071 | 0.64 | 0.691 | 0.331 | CNN |
| 0.735 | 0.149 | 0.669 | 0.701 | 0.914 | 0.086 | 0.704 | 0.633 | 0.338 | RF |
| 0.694 | 0.162 | 0.679 | 0.776 | 0.905 | 0.095 | 0.785 | 0.574 | 0.359 | kNN |

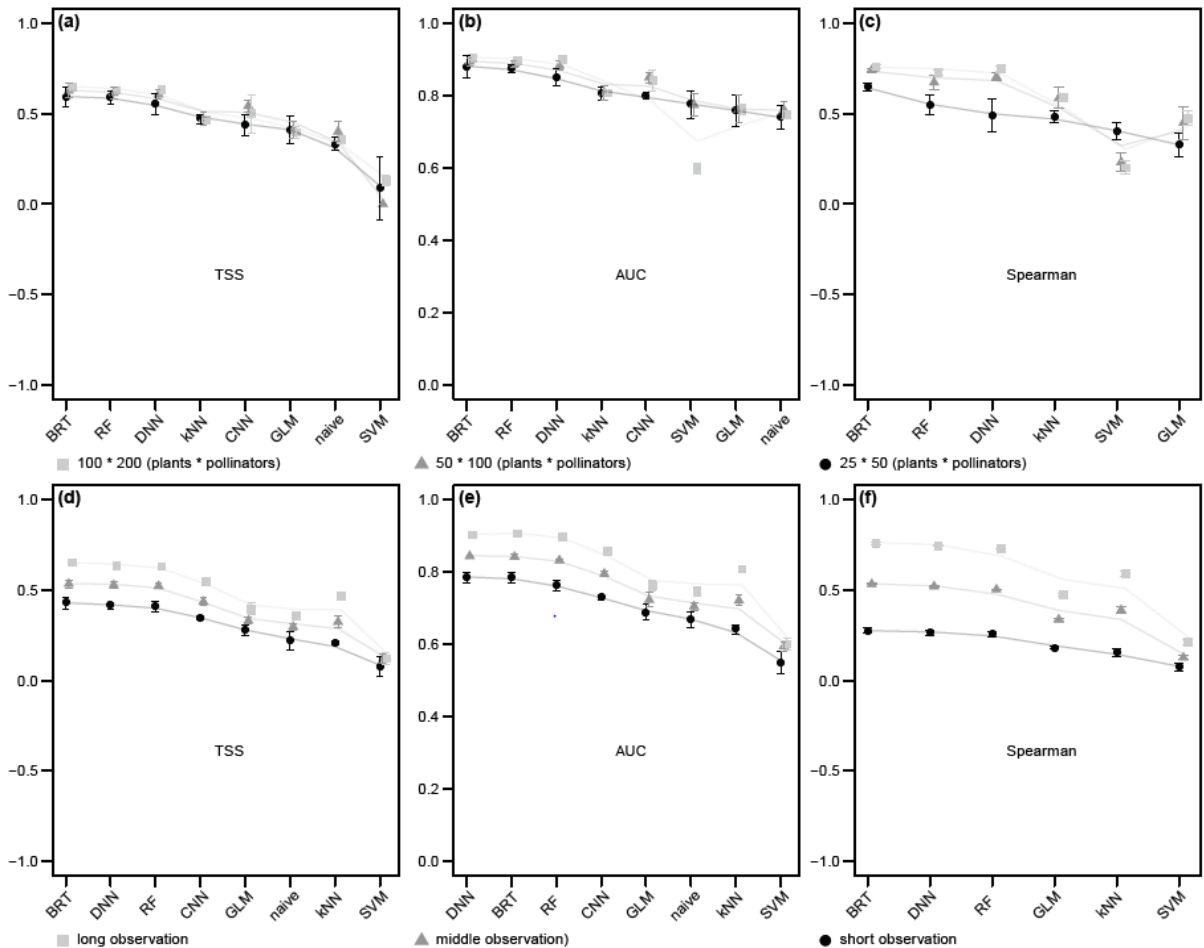

**Figure S2**: Predictive comparison of machine learning models for varying network sizes (a-c)

and varying observation times (d-f). We compared three network sizes (20*50, 50*100, and 100*200 species*species), for presence-absence plant-pollinator interactions (a, b) and plant-pollinator interaction counts , (c) and three observation times (0.007, 0.0032, 0.12) for presence-absence plant-pollinator interactions (d, e) and plant-pollinator interaction counts (f). We used TSS (true skill statistic) and AUC (area under the curve) for estimating predictive performance for presence-absence (a, b, d, e) and Spearman Rho correlation factor for count frequencies (c, f).

**Table S5:** Results for baseline models. TSS = True skill statistic. AUC and TSS for presence-absence plant-pollinator interaction models and Spearman Rho correlation factor for plant-pollinator interaction count models.

| Measure | dnn | cnn | knn | naive | RF | boost | glm | SVM | glm_step |
|---|---|---|---|---|---|---|---|---|---|
| AUC | 0.5 | 0.51 | 0.51 | 0.51 | 0.52 | 0.51 | 0.51 | 0.51 | 0.51 |
| TSS | 0 | 0 | 0.01 | 0 | 0 | 0.01 | 0 | 0 | 0.01 |
| Spearman | -0.02 | - | -0.01 | - | -0.03 | -0.02 | -0.03 | -0.04 | -0.03 |

**Table S6:** Decrease in percent of performance from training to testing (generalization error) for varying network sizes. AUC for presence-absence plant-pollinator models and Spearman Rho correlation factor for plant-pollinator interaction count models.

| NetworkSize | DeacreaseInPercent | dnn | cnn | knn | naive | RF | boost | glm | SVM | glm_step |
|---|---|---|---|---|---|---|---|---|---|---|
| 20*50 | AUC | 11 | 10.4 | 17.4 | 7.3 | 11.8 | 9.3 | 13 | 7 | 6 |
| 50*100 | AUC | 6.1 | 5.2 | 17 | 2.6 | 7 | 5.8 | 8.1 | 1 | 2.7 |
| 100*200 | AUC | 1.5 | 2.5 | 15.1 | 1.7 | 6.5 | 3.1 | 4.2 | 1.1 | 2.2 |
| 20*50 | Spearman | 36.7 | - | 37.1 | - | 33.4 | 20.9 | 40.1 | -11.9 | 19.9 |
| 50*100 | Spearman | 11.9 | - | 28.6 | - | 21.4 | 9.7 | 19.6 | -7.9 | 8.8 |
| 100*200 | Spearman | 2.6 | - | 26 | - | 14.5 | 4.7 | 9 | -8.7 | 6.2 |

## Comparison of causal inferential performance

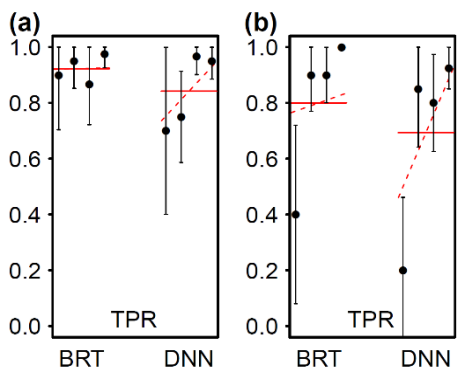

**Figure S3:** Averaged true positive rates for causal inferential performance on a 50*100 simulated plant-pollinator network. We tested the performance for DNN and BRT on one to four true trait-trait interactions. Red line is the averaged mean of true positive rates. Results were higher for species presence-absence (a) interactions than for species interaction counts (b).

Hummingbird-plant networks

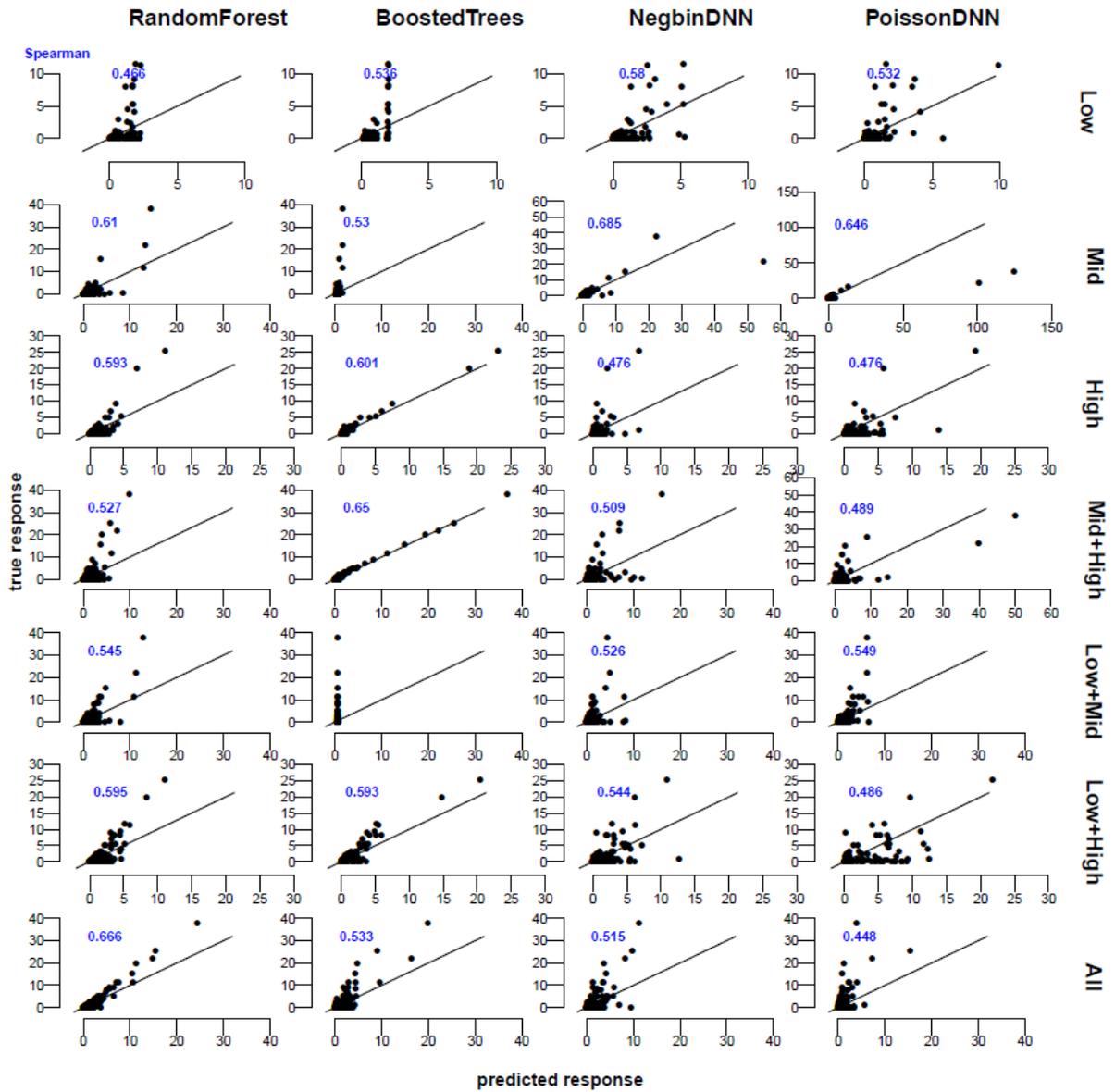

**Figure S4:** For estimating the predictive power of random forest, BRT, DNN with a negative binomial log-likelihood, and a DNN with a Poisson log-likelihood, we plotted the true observations versus the predictions. BRT and random forest showed best fits. Spearman rho correlation was used to quantify the predictive power and therefor the fit.

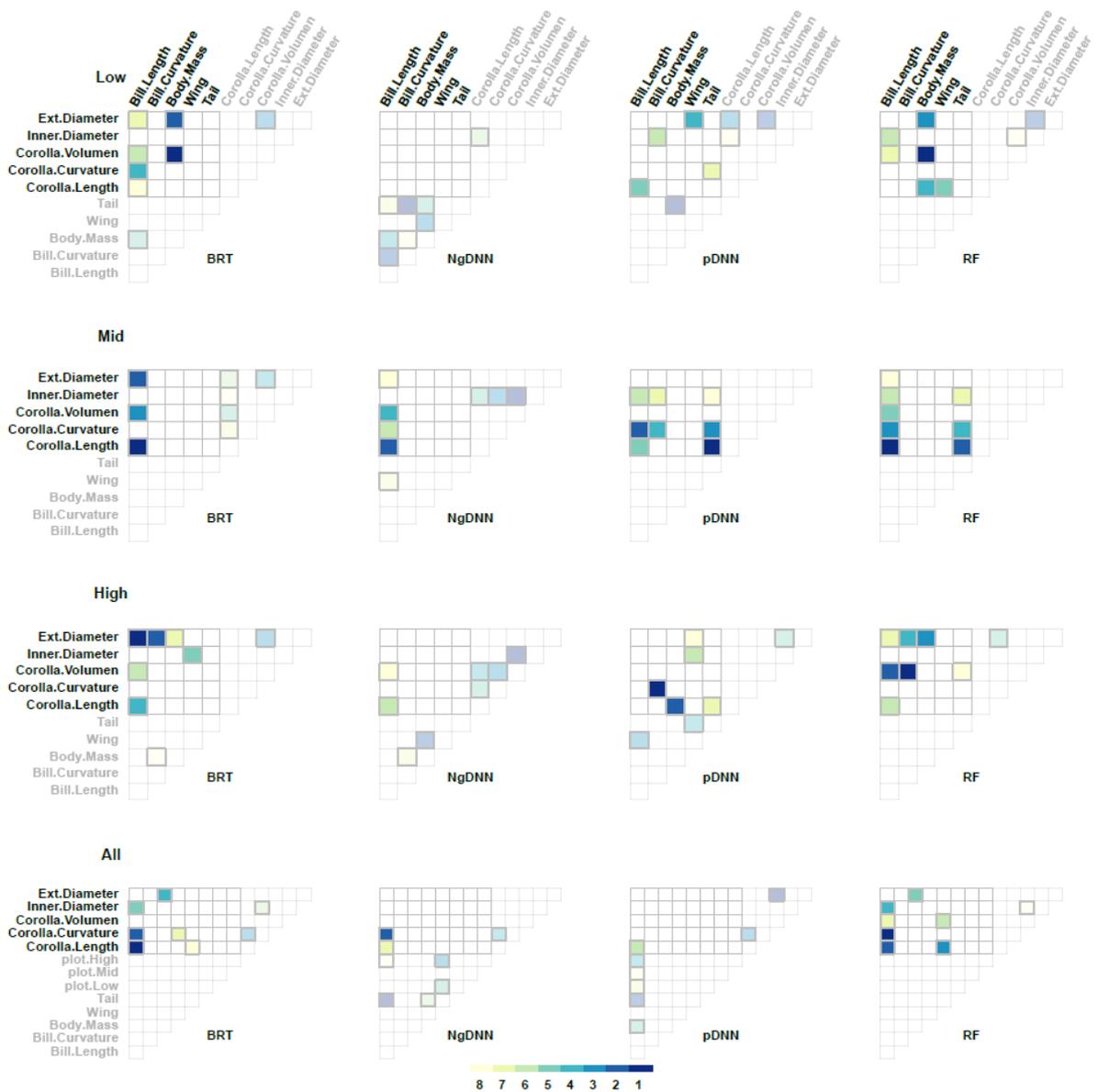

**Figure S5:** Low network - BRT, DNN with poisson log-likelihood, DNN with negative binomial log-likelihood, and random forest were fit to the low plant-hummingbird network. The four traits with highest interaction strength versus all traits and for each of those the top two pairwise interactions were here visualized. Random forest identified mainly trait combinations between plants and hummingbirds with higher interaction strengths.